\documentstyle[12pt,epsfig]{article}

\begin{document}
\baselineskip=23pt

\vspace{1.2cm}

\begin{center}
{\Large \bf  Dark Energy Density in Brane World}

\bigskip

Hai-Bao Wen\footnote{Email address: wenhb@mail.ihep.ac.cn}
\\
{\em Institute of High Energy Physics,
Chinese Academy of Sciences} \\
{\em P.O.Box 918(4), 100049 Beijing, China}
\\
Xin-Bing Huang\footnote{Email address: huangxb@pku.edu.cn,
huangxb@mail.phy.pku.edu.cn}
\\
{\em Department of Physics, Peking University,
 100871 Beijing, China}

\end{center}

\bigskip

\centerline{\large Abstract}

We present a possible explanation to the tiny positive
cosmological constant under the frame of AdS$_5$ spacetime
embedded by a dS$_4$ brane. We calculate the dark energy density
by summing the zero point energy of massive scalar fields in
AdS$_5$ spacetime. Under the assumption that the radius of AdS$_5$
spacetime is of the same magnitude as the radius of observable
universe, the dark energy density in dS$_4$ brane is obtained,
which is smaller than the observational value. The reasons are
also discussed.

\vspace{1.2cm}


\vspace{1.2cm}

\newpage

\section{Introduction}\hspace*{\parindent}
In Randall-Sundrum models,\cite{rs99} the vacuum energy in the
bulk is given by a bulk cosmological constant, which can not be
calculated from more fundamental theory, such as quantum field
theory. It is well known that there are several kinds of scalar
fields in vacuum in modern physics. One of the most famous
examples is the inflaton field, which drives the inflation of
universe. In this Letter, we therefore study the vacuum energy
density from the viewpoint of the zero point energy of scalar
fields.

Recent astronomical observations on Type Ia
supernova~\cite{supernova,rie04} and the cosmological microwave
background radiation~\cite{cmb,ben03} indicate that our universe
is spatially flat and accelerating at present, which supports for
a concordant cosmological model of inflation + cold dark matter +
dark energy. Dark energy is one of the most popular topics in
recent years, which has invoked much discussion in both
theoretical physicists~\cite{lee04} and astronomers. Dark energy
can be looked as vacuum energy from the cosmologists' point of
view. The vacuum energy density calculated by quantum field theory
is much larger than the possible observed value.\cite{wei,cpt92}
Many scenarios have been proposed in trying to solve this problem.
However, up to now, there has not yet been a generally accepted
theory that can give a vacuum energy density, whose order is the
same as that of the observed value.

Here we propose a scenario to study this problem under the
framework of brane world. Following Randall and
Sundrum,\cite{rs99} we consider a dS$_4$ brane embedded in an
AdS$_5$ spacetime. The solution of the Einstein equation and the
localization of gravity in this setup have been
acquired.\cite{local,kt02} On the other hand, Chang {\em
et~al.}~\cite{cgg03} gave the exact solutions of massive scalar
fields in AdS$_n$ manifold, where the discrete spectra of the mass
eigenvalues were acquired. Therefore following the method proposed
by Chang and Huang,\cite{ch02} we assume that the total vacuum
energy in the bulk is given by the sum of the zero point energy of
massive scalar fields. Hence the observable effective
four-dimensional vacuum energy density in brane should be the
five-dimensional vacuum energy density times the radius of the
extra dimension.

In supersymmetric grand unified theories (GUT), for instance, in
supersymmetric SO(10) model, all of the gauge field coupling
constants can be unified. To realize the unification of
gravitation with other gauge fields, the fundamental energy scale
of five-dimensional gravity should be the same as the
supersymmetric GUT energy scale. When we select the supersymmetric
GUT energy scale $M_{G}$ as the fundamental energy scale of
five-dimensional gravity, simultaneously the radius of the
AdS$_{5}$ space-time is admitted by the astronomical observations,
the tiny but nonzero four-dimensional cosmological constant can be
acquired, which is slightly smaller than the value given by modern
astronomical observations.\cite{rie04,cmb} The possible reasons
leading to this smallness are also discussed at the end of this
paper.

\section{Basic Setup}
\hspace*{\parindent}Considering an AdS brane world with a dS$_4$
brane embedded at $y=0$. It is assumed that there are gravity and
vacuum energy in the bulk. Hence, the action can be written as
\begin{equation}
\label{wen1} {\cal S}=\int
d^4x~dy\left[\sqrt{G}\left(\frac{M^3}{2}R-\Lambda
\right)-\sqrt{-g}V\delta(y)\right]~,
\end{equation}
where $\displaystyle\frac{\Lambda}{M^{3}}$ ($\Lambda<0$) and $M$
are respectively the bulk cosmological constant and fundamental
energy scale of five-dimensional gravity. According to the
assumption, the metric on the brane world can be written as
follows (here five-dimensional suffices are denoted by capital
Latin and the four-dimensional suffices by the Greek ones),
\begin{equation}
\label{wen2}
 ds^2=e^{2A(y)}g_{\mu\nu}dx^\mu
dx^\nu - dy^2~,~~~~~~\mu,\nu=0,1,2,3,
\end{equation}
where $g_{\mu\nu}$ is the metric on the brane and
\begin{equation}
\label{wen3} g_{\mu\nu}={\rm diag}\left(1,~ -e^{2 \sqrt{\lambda}
x_{0}},~ -e^{2 \sqrt{\lambda} x_{0} },~ -e^{2 \sqrt{\lambda} x_{0}
}\right)~.
\end{equation}
Here $\sqrt{\lambda}$ ($\lambda>0$) is the curvature of the dS
brane. According to de Sitter~\cite{wei}, $\lambda_{4}=3\lambda$
is the four-dimensional cosmological constant of the brane.

The five-dimensional Einstein equations for the above action read
\begin{equation}
\label{wen4} R_{MN}-\frac{1}{2}G_{MN}R= \frac{1}{M^3}\left[\Lambda
G_{MN}-\frac{\sqrt{-g}}{\sqrt{G}}Vg_{\mu\nu}\delta^\mu_M\delta^\nu_N\delta(y)
\right]~.
\end{equation}
By solving the Einstein equations of this setup, Warp factor
$A(y)$ is acquired as follows,\cite{local,warp}
\begin{eqnarray}
\label{wen8} A(y)=\log(L\sqrt{\lambda}\sinh\frac{c-|y|}{L}),
\end{eqnarray}
where $c$ is a positive integration constant that represents the
distance between the brane and the horizon. The radius of the bulk
is given by
\begin{eqnarray}
\label{wen9} L=\sqrt{\frac{6M^{3}}{-\Lambda}}\,.
\end{eqnarray}
Imposing the normalization condition $A(0)=0$ on
Eq.\,(\ref{wen8}), we can write $\lambda$ in terms of $c$, namely
\begin{equation}
\label{wen10} \lambda=L^{-2}\sinh^{-2}\frac{c}{L}.
\end{equation}
Due to the presence of brane at $y=0$, the first order
differential of $A(y)$ is not continuous. Fine-tuning leads to the
relation between the brane tension $V$ and the distance $c$. The
brane tension is expressed by
\begin{equation}
\label{wen11} V=\frac{6M^{3}}{L}\coth\frac{c}{L}.
\end{equation}
The localization of gravity has been studied by several
Authors.\cite{local} Their results show that the gravity in
AdS$_5$ space-time with a dS$_4$ brane is localized. Moreover,
Kehagias and Tamvakis indicated that the conventional Newton law
is valid in this setup.\cite{kt02}


We now present the derivation of the four-dimensional effective
Planck energy scale $M_{P}$ ($M_{P}$ is defined by
$M_{P}^{2}=\frac{1}{8\pi G_{N }}$ with $G_{N}$ being the Newtonian
gravitational constant). The zero mode of four-dimensional
graviton following from the solutions (\ref{wen2}) and
(\ref{wen8}) is described by an effective action, which is derived
from Eq.\,(\ref{wen1}),
\begin{equation}
\label{huang1} {\cal S}_{eff} \supset \frac{M^{3}}{2} \int
d^{4}x\int^{c}_{-c}dy~e^{2A(y)}\sqrt{-g}~\overline{R},
\end{equation}
where $\overline{R}$ denotes the four-dimensional Ricci scalar
made out of $g_{\mu\nu}(x)$, in contrast to the five-dimensional
Ricci scalar $R$, which is made out of $G_{MN}(x,y)$. Because the
effective graviton field is four-dimensional, we can explicitly
perform the $y$ integral to acquire a purely four-dimensional
action. From this we obtain

\begin{equation}
\label{huang2} M_{P}^{2}=M^3\int^c_{-c}e^{2A(y)} dy=\frac{\lambda
M^{3}L^{3}}{2}(\sinh \frac{2c}{L}-\frac{2c}{L})~.
\end{equation}

Here the simple relation between the four-dimensional effective
Planck energy scale $M_{P}$ and the five-dimensional fundamental
scale of gravity $M$ is acquired, which is quite different from
the relations given by Arkani-Hamed {\it et al.}\cite{add98} or
Randall-Sundrum models.

\hspace*{\parindent}In Randall-Sundrum model 2, a flat brane was
embedded in an AdS$_5$ bulk, there was a fine-tuning between the
bulk cosmological constant and the brane tension. Now, there is
not a necessary fine-tuning between them in our setup. How can we
determine the bulk cosmological constant? We will connect the bulk
cosmological constant with the bulk energy density, and
furthermore calculate the bulk energy density by summing up the
zero-point energies of massive scalar field in the bulk. In
AdS$_4$ space-time, the same idea has been carried out and an
interesting result has been presented.\cite{ch02}

To discuss the massive scalar field in our brane world, we
introduce other coordinates to describe the brane world. It will
be shown that those two coordinates are equivalent. AdS$_5$
space-time can be treated as a submanifold of a pseudo-Euclidean
six-dimensional space with Cartesian coordinates $\xi^a$
\begin{eqnarray}
\label{wen12}
(\xi^0)^2-(\xi^1)^2-(\xi^2)^2-(\xi^3)^2-(\xi^4)^2+(\xi^5)^2=L^2~,
\nonumber\\
ds^2=(d\xi^0)^2-(d\xi^1)^2-(d\xi^2)^2-(d\xi^3)^2-(d\xi^4)^2+(d\xi^5)^2.
\end{eqnarray}
The brane is a hypersurface in this coordinates, which can be
expressed as
\begin{equation}
\label{wen121}  \left\{\begin{array}{l}
\displaystyle(\xi^0)^2-(\xi^1)^2-(\xi^2)^2-(\xi^3)^2-(\xi^4)^2=-\frac{1}{\lambda},\\[0.2cm]
\displaystyle\xi^{5}=\sqrt{L^{2}+\frac{1}{\lambda}}.\end{array}
\right.
\end{equation}
Obviously the symmetry group of AdS$_5$ is the conformal group
$SO(4,2)$. Introduce the Beltrami coordinates
\begin{equation}
\label{wen13} z^{i}=L\frac{\xi^{i}}{\xi^{5}}~,~~~~~~i=1,2,3,4.
\end{equation}
Chang {\em et al.}\cite{cgg03} indicated that in the coordinate
$(\xi^0,z^\alpha)$, the metric of AdS$_5$ can be written as
\begin{equation}
\label{wen14}
  ds^2=\frac{1}{1-L^{-2}\xi^0\xi^0}d\xi^0d\xi^0-(1-L^{-2}\xi^0\xi^0)
  \frac{d{\bf z}(I-L^{-2}{\bf z}'{\bf z})^{-1}d{\bf z}'}
      {1-L^{-2}{\bf z}{\bf z}'},
\end{equation}
where the vector ${\bf z}$ denotes $(z^1,~z^2,~z^3,~z^4)$ and
${\bf z}'$ represents the transpose of the vector ${\bf z}$.

In the spherical coordinate $(z^1,~z^2,~z^3,~z^4)\longrightarrow
(\rho,~\theta,~\phi,~\omega)$, the $SO(4,2)$ invariant metric
(\ref{wen14}) is in the form
\begin{eqnarray}
\label{wen15} ds^2 &=& \frac{1}{1-L^{-2}\xi^0\xi^0}d\xi^0d\xi^0
-(1-L^{-2}\xi^0\xi^0) \nonumber
\\
 & &\times\left[\frac{d\rho^2}{\left(1-\frac{\rho^2}{L^{2}}\right)^2}
     +\frac{\rho^2}{1-\frac{\rho^2}{L^{2}}} \left(d\theta^2+\sin^2\theta
d\phi^2+\sin^2\theta \sin^2\phi d\omega^2\right) \right],
\end{eqnarray}
and the brane can be written as
\begin{equation}
\label{wen151} \xi^0\xi^0-\rho^{2}\left(1+\frac{1}{\lambda
L^{2}}\right)=-\frac{1}{\lambda}.
\end{equation}
There is a horizon\cite{cgg03} in the coordinate $(\xi^0,{\bf
z})$. Comparing with the metric in Eqs.\,(\ref{wen2}) and
(\ref{wen3}), we limit our physical brane world in the region of
$\vert\xi^0\vert<L$. In this region of AdS, we can introduce a
time-like variable $\tau$ as
\begin{equation}
\label{wen16} \zeta \equiv \frac{\xi^0}{L}\equiv
\sin(\frac{\tau}{L}).
\end{equation}
Then we have a Robertson-Walker-like metric
\begin{eqnarray}
\label{wen17} ds^2=d\tau^2-R^2(\tau)\left[\frac{d\rho^2}
{\left(1-\frac{\rho^2}{L^{2}}\right)^{2}}
+\frac{\rho^2}{1-\frac{\rho^2}{L^{2}}} \left(
d\theta^2+\sin^2\theta ~d\phi^2+\sin^2\theta~ \sin^2\phi ~
d\omega^2 \right) \right],
\end{eqnarray}
where we have used the notation
$R(\tau)=\displaystyle\cos(\frac{\tau}{L})$. The dS$_4$ brane thus
becomes
\begin{equation}
\label{wen171} \frac{\cos^{2}\frac{\tau}{L}}{1+\frac{1}{\lambda
L^{2}}}+\frac{\rho^{2}}{L^{2}}=1.
\end{equation}
It can be checked that the metrics (\ref{wen2}) and (\ref{wen17})
describe the same bulk manifold, in which $L$ is the radius of
AdS$_{5}$ space-time. It is obvious that Eq.\,(\ref{wen171}) gives
a dS surface, whose curvature is $\sqrt{\lambda}$. Then, one can
have a conclusion that those two coordinates are equivalent and
describe the same brane world.

To study dark energy density in the bulk, we introduce scalar
field in this space-time. The equation of motion for a massive
scalar field in AdS$_{5}$ space-time is of the form
\begin{equation}
\label{wen18}
\begin{array}{l}
\displaystyle\left[\frac{1}{R^4}\frac{\partial}{\partial\tau}
\left(R^4\frac{\partial}
{\partial\tau}\right)-\frac{(1-L^{-2}\rho^2)^{\frac{5}{2}}}{R^2\rho^3}
\frac{\partial}{\partial\rho}
\left(\frac{\rho^3}{(1-L^{-2}\rho^2)^\frac{1}{2}}\frac{\partial}{\partial
\rho}\right)\right.\\[1cm]
\displaystyle~~~~~~~~~~~\left.-\frac{1-L^{-2}\rho^2}{R^2\rho^2\sin^2\theta}
\left(\frac{\partial}
{\partial\theta}\left(\sin^2\theta\frac{\partial}{\partial\theta}\right)+
\frac{1}{\sin\phi}\frac{\partial}
{\partial\phi}\left(\sin\phi\frac{\partial}{\partial\phi}\right)+
\frac{1}{\sin^2\phi}\frac{\partial^2}{\partial\omega^2}\right)\right.\\[1cm]
\displaystyle~~~~~~~~~~~\left.
+m_{0}^2\right]\Phi(\tau;\rho,\theta,\phi,\omega)=0.
\end{array}
\end{equation}
In the literature,\cite{cgg03,ch02,lg82,fro74} the solutions of
Eq.\,(\ref{wen18}) were given by variable separation as follows,
\begin{eqnarray}
\label{wen19} \Phi(\tau;\rho,\theta,\phi,\omega)&=&T(\tau)U(\rho)
Y_{lmn}(\theta,\phi,\omega).
\end{eqnarray}
The reduced equations of motion in terms of $T(\tau)$, $U(\rho)$
and $Y_{lmn}(\theta,\phi,\omega)$ are of the form
\begin{equation}
\label{radiusdif} \frac{\partial^2 U}{{\partial\rho}^2}+
\frac{-2L^{-2}\rho^2+3}{\rho(1-L^{-2}\rho^2)}\frac{\partial
U}{\partial\rho}+\frac{k^2}{(1-L^{-2}\rho^2)^2}U-
\frac{l(l+2)}{\rho^2 (1-L^{-2}\rho^2)}U =0,\nonumber
\end{equation}
\begin{equation}
\label{timedif}
R^2\frac{d^2T}{{d\tau}^2}+4R\frac{dR}{d\tau}\frac{dT}{d\tau}
+(m_{0}^2R^2+k^2)T = 0,
\end{equation}
\begin{eqnarray}
\label{anglardif} \frac{\partial^2Y_{lmn}}{{\partial\theta}^2}&+&
2{\cot}\theta\frac {\partial Y_{lmn}}{\partial\theta}
+\frac{1}{\sin^2\theta}\frac{\partial^2Y_{lmn}}{{\partial\phi}^2}+
\frac{1}{\sin^2\theta}{\cot}\phi\frac {\partial
Y_{lmn}}{\partial\phi} \nonumber \\
& +&\frac{1}{{\rm sin}^2\theta}\frac{1}{{\rm
sin}^2\phi}\frac{\partial^2Y_{lmn}}
{{\partial\omega}^2}+l(l+2)Y_{lmn}=0.
\end{eqnarray}
It is obvious that the solutions of the angular part are of the
three-dimensional spherical harmonic functions
$Y_{lmn}(\theta,\phi,\omega)$.

For convenience, setting $\sigma=L^{-1}\rho$, we rewrite
Eq.\,(\ref{radiusdif}) as
\begin{equation}
\label{radiousre}\frac{\partial^2 U}{{\partial\sigma}^2}+\left[
\frac{2}{\sigma}+\frac{1}{\sigma(1-\sigma^2)}\right]
\frac{\partial U}{\partial\sigma}+\frac{k^2L^2}{(1-\sigma^2)^2}U
-\frac{l(l+2)}{\sigma^2 (1-\sigma^2)}U =0.
\end{equation}
The solution of Eq.\,(\ref{radiousre}) is of the form~\cite{cgg03}
\begin{equation}
\label{radius}\displaystyle U(\rho)={\mathcal
C}(\frac{\rho}{L})^l(1-\frac{\rho^2}{L^2})^{\frac{\mu}{2}}
{\mathcal
F}(\frac{1}{2}(l+\mu+1),\frac{1}{2}(l+\mu),l+2;\frac{\rho^2}{L^2}),
\end{equation}
where ${\mathcal C}$ is the normalization constant, ${\mathcal F}$
is the confluent hypergeometric function, and $\mu$ satisfies
\begin{equation}
\label{mu} \mu^2-3\mu+k^2L^2=0.
\end{equation}

The time-like evolution equation (\ref{timedif}) can be
transformed into the associated Legendre equation, therefore the
solutions are presented as
\begin{eqnarray}
\label{wen20} T_{1}(\tau) \propto \frac{1}{\cos\frac{\tau}{L}}
{\mathcal P}_{I}^{N} \left(\zeta\right)~,& &~~~~~~T_{2}(\tau)
\propto \frac{1}{\cos\frac{\tau}{L}} {\mathcal Q}_I^N
\left(\zeta\right)~,\nonumber\\
I=-\frac{1}{2}+\sqrt{4+m_{0}^2L^2},& &
~~~~~~N={\sqrt{\frac{9}{4}-k^2L^2}},
\end{eqnarray}
where ${\mathcal P}_I^N(\zeta)$ and ${\mathcal Q}_I^N(\zeta)$ are
the associated Legendre functions. Because ${\mathcal
Q}_I^N(\zeta)$ becomes infinite on the boundary
$\vert\xi^{0}\vert=L$, we would ignore ${\mathcal Q}_I^N(\zeta)$.
The natural boundary condition of ${\mathcal P}_I^N(\zeta)$ on
$\vert\xi^{0}\vert=L$ requires $I,~ N$ to be integers. From
Eq.\,(\ref{wen20}), we can acquire the discrete spectrum of the
eigen mass and wave number of the scalar field as follows,
\begin{equation}
\label{mass}
\begin{array}{l}
L^2m_{0}^2+\frac{15}{4}=I(I+1),\\[0.4cm]
-k^2L^2+\frac{9}{4}=N^2~,\quad  |N|\leq{I}.
\end{array}
\end{equation}
Because of orthogonality of the associated Legendre functions
${\mathcal P}_{I}^{N}(\zeta)$ and the spherical harmonic functions
$Y_{lmn}$, we can write the orthogonal wave functions of scalar
fields in AdS$_5$ space-time in the form
\begin{equation}\label{solution}
\begin{array}{l}
\Phi_{NIlmn}(\tau;\rho,\theta,\phi,\omega)\propto U (\rho){(\cos
\frac{\tau}{L})}^{-1} {\mathcal P}_I^N\left(
\zeta\right)Y_{lmn}(\theta,\phi,\omega).
\end{array}
\end{equation}

In quantum field theory, the vacuum is treated as the ground state
of quantum fields. Then the vacuum energy should be the sum of the
zero point energy of all kinds of quantum fields. Similarly, we
assume that the bulk energy density comes from the zero point
energy of all kinds of fields in the bulk. We can sum up the
contributions of scalar fields with different mass since the
discrete mass spectrum has been obtained for scalar fields in
AdS$_5$ space-time. Einstein mass-energy formula in AdS$_5$ is of
the form (with $\hbar=c=1$) similar to that in Minkowski
space-time. Because angular momentum must be zero, the ground
state energy of a simple harmonic oscillator is
$E^{2}=m_{0}^{2}+k^{2}$. Thus the energy density of AdS vacuum
devoted by scalar fields can be presented as follows,
\begin{eqnarray}
\langle\rho\rangle_{s}=\pi^2\sum_{m_{0}}\sum_{k}\frac{k^3}{{(2\pi)}^4}\delta
k\sqrt{m_{0}^2+k^2},
\end{eqnarray}
where $\delta k$ is the wave-vector difference of two eigen
states. When $\displaystyle N=0$, $\displaystyle k=\frac{3}{2L}$
and $N=\pm 1$, $\displaystyle k=\frac{\sqrt5}{2L}$. Therefore,
Eq.\,(\ref{mass}) can be used to obtain the energy density of
vacuum as follows
\begin{equation}\label{sum1}
\langle\rho\rangle_{s}=\frac{3-\sqrt{5}}{32\pi^{2}
L^{5}}\sum_{I=I_{\rm min}}^{I_{\rm max}} \left[
\frac{5\sqrt{5}}{4}\sqrt{I(I+1)-\frac{5}{2}}
+\frac{27}{8}\sqrt{I(I+1)-\frac{3}{2}}\right],
\end{equation}
where $I_{\rm min}=3$, obained from the first equation of
(\ref{mass}), and  $I_{\rm max}$ is the cutoff of mass spectrum.
We would estimate $I_{\rm max}$ as the five-dimensional
fundamental energy scale $M$. Namely, the maximal energy $E_{\rm
max}$ is the energy corresponding to the cutoff $I_{\rm max}$
\begin{eqnarray}
\label{max} E_{\rm max}=M=\frac{1}{L}\sqrt{I_{\rm max}(I_{\rm max}
+1)-\frac{3}{2}}.
\end{eqnarray}
Substituting Eq.\,(\ref{max}) into Eq.\,(\ref{sum1}) yields
\begin{equation}\label{sum2}
\displaystyle\langle\rho\rangle_{s}
=\frac{3-\sqrt{5}}{32\pi^{2}}(\frac{5\sqrt{5}}{8}+\frac{27}{16})\frac{M^2}{L^3}
=\frac{0.074}{\pi^2}\frac{M^2}{L^3}.
\end{equation}

We therefore obtain the vacuum energy density in the bulk by
adding the zero point energy of scalar fields. It is reasonable to
assume that the vacuum energy density of the brane is of the
magnitude of the vacuum energy density in the bulk
$\langle\rho\rangle_{s}$ times the radius of extra dimension $L$,
namely
\begin{equation}
\label{sum3} \displaystyle\langle\rho\rangle_4
\sim\displaystyle\langle\rho\rangle_sL
=\frac{0.074}{\pi^2}\frac{M^2}{L^2}.
\end{equation}
Having argued in the present introduction, we select the
supersymmetric GUT energy scale $M_{G} = 3.1 \times 10^{16} {\rm
GeV}$ as the fundamental energy scale of five-dimensional gravity
$M$. Then Eq.\,(\ref{sum3}) gives
\begin{equation}
\label{sum4} \displaystyle
L\sim\frac{M_{G}}{\pi}\sqrt{\frac{0.074}{\langle\rho\rangle_4}}
=\frac{M_{G}}{\pi}\sqrt{8\pi
G_N\frac{0.074}{\lambda_4}}=\frac{M_{G}}{\pi M_{p}}\sqrt{
\frac{0.074}{\lambda_4}}.
\end{equation}

Obviously, one can reasonably believe that the radius of AdS$_{5}$
space-time should be of the same magnitude as the radius of our
observable universe, that is,
\begin{equation}
\label{magnitude} L\sim\frac{1}{H_{0}}\sim 10^{26}~\rm{m},
\end{equation}
where $H_{0}$ is the present Hubble constant. Then the
four-dimensional cosmological constant $\lambda_4$ is
\begin{equation}
\label{Lsqlambda} \lambda_4\sim
10^{-58}~\textrm{m}^{-2}~(10^{-90}~\textrm{GeV}^{2}).
\end{equation}
The recent cosmological observations\cite{rie04,cmb,ben03}
indicated that the cosmological constant is about $1.0 \times
10^{-52}~\textrm{m}^{-2}$. Hence the four-dimensional cosmological
constant we obtained is smaller than the observational value. The
reasons, we think, mainly include the two aspects: (1) The zero
point energy of other scalar fields at the brane, i.e., Higgs
field, should also give benefit to the vacuum energy. (2) Since
there are no experimental implications that the supersymmetric GUT
or other GUT energy scale can not be higher than $M_{G}$ till now,
it is possible that the GUT energy scale is higher than $M_{G}$.

In conclusion, the unification of gravity with other gauge field
interactions and a tiny but nonzero dark energy density are two
fundamental problems in modern physics. Various attempts have been
made in trying to solve them. It can be found in this study that
in the model of an AdS$_5$ space-time embedded by dS$_4$ brane,
the hierarchy problem and the cosmological constant problem could
be correlated and explained simultaneously. The dark energy
density in the bulk is calculated by summing the zero-point
energies of massive scalar fields in AdS$_5$ space-time. Since the
radius of AdS$_5$ spacetime is of the same magnitude as the radius
of observable universe, the dark energy density in dS$_4$ brane is
obtained, which is slightly smaller than the observational value.
The reasons are also discussed in this study.

We would like to thank Professor Z. Chang for his enlightening
discussion.

\end{document}